# Intrinsic Defect Energetics and Fluorine Doping Effects in $Li_2CO_3$ and $Li_2O_2$: A First-Principles Study

*Youjeong Choi*[1], *Tasuku Sugiura*[2], *Keisuke Mukai*[3], *Nanako Ishihara*[4], *Shuji Nakanishi*[4,5], *and Teruyasu Mizoguchi*[1,2,*]

[1] Department of Materials Engineering, The University of Tokyo, Tokyo 113-8656, Japan

[2] Institute of Industrial Science, The University of Tokyo, Tokyo 153-8505, Japan

[3] National Institute for Fusion Science, Gifu 509-5292, Japan

[4] Research Center for Solar Energy Chemistry, Graduate School of Engineering Science, The University of Osaka, Osaka 560-8531, Japan

[5] Innovative Catalysis Science Division, Institute for Open Transdisciplinary Research Initiatives, The University of Osaka, Osaka 565-0871, Japan

*Corresponding author. E-mail: teru@iis.u-tokyo.ac.jp

## ABSTRACT

Lithium carbonate, $Li_2CO_3$, is a thermodynamically stable carbonate phase whose defect energetics are closely related to its stability and decomposition behavior in various lithium-based electrochemical systems. These properties of $Li_2CO_3$ are particularly important in lithium–oxygen battery environments. In these systems, $Li_2CO_3$ can form as a parasitic discharge product alongside $Li_2O_2$, the primary discharge product, leading to performance degradation. However, compared with $Li_2O_2$, the intrinsic defect thermodynamics of $Li_2CO_3$ and how chemical doping modifies its defect energetics remain insufficiently understood. In this study, first-principles calculations were performed to systematically analyze the intrinsic point-defect energetics of $Li_2CO_3$ and to evaluate the effects of fluorine doping on vacancy formation energies in $Li_2CO_3$ and $Li_2O_2$. Intrinsic defect analysis reveals that defect behavior is predominantly governed by lithium-related defects. Upon fluorine doping, the carbon vacancy formation energies decrease selectively with destabilizing the carbonate framework. Lithium vacancy formation energies decrease partially in $Li_2CO_3$ under certain conditions, while a reduction in neutral lithium vacancy formation energy is observed in $Li_2O_2$. These results suggest that fluorine doping modulates the defect energetics of both discharge products, potentially providing a thermodynamic basis for controlling the stability of $Li_2CO_3$ and $Li_2O_2$ under thermodynamic conditions representative of lithium–oxygen batteries.

## INTRODUCTION

Lithium carbonate ($Li_2CO_3$) is a representative carbonate phase in various lithium-based electrochemical systems. It appears as an inorganic component of the solid-electrolyte interphase in lithium–ion batteries, as a major discharge product in lithium–carbon dioxide batteries, and as a parasitic discharge product in lithium–oxygen batteries (LOBs).[1–9] Due to the robust bonding network of carbonate units, $Li_2CO_3$ is thermodynamically stable and exhibits poor ionic and electronic conductivities.[7] Because these characteristics make $Li_2CO_3$ difficult to electrochemically decompose once formed, controlling its formation and accumulation is essential for improving the performance of these electrochemical systems.

To regulate the formation and accumulation of discharge products including $Li_2CO_3$, various cathode modification strategies have been investigated in LOBs and lithium–carbon dioxide batteries. For example, binder-free cathodes composed of conductive metal–organic framework (MOF) nanorods supported on biomass-derived porous carbon enhanced the reversible generation and decomposition of $Li_2CO_3$, and improved long-term cycling stability in lithium–carbon dioxide batteries.[10,11] In LOBs, a cathode catalyst containing atomically dispersed Ta on $Co_3O_4$ was reported to suppress side reactions and mitigate the formation of $Li_2CO_3$ byproducts.[12] These electrode-level modification strategies provide an effective means of regulating $Li_2CO_3$.

Along with these electrode-level approaches, atomic-level investigation of $Li_2CO_3$ is crucial for clarifying the intrinsic factors governing their thermodynamic stability and charge transport properties. These properties are closely related to defect formation and stability, and elucidating defect energetics provides a foundation for understanding the intrinsic properties of $Li_2CO_3$. This atomic-scale insight is particularly important in LOBs, where $Li_2CO_3$ forms alongside $Li_2O_2$ as a discharge product. [13–19]

In LOBs, the accumulation of $Li_2CO_3$ on the cathode impedes charge transport and requires high charging overpotentials for decomposition, leading to severe losses in the energy efficiency and cyclability of LOBs.[6,8,20,21] To enhance charge transport and decomposability within the discharge products of LOBs, doping strategies have been actively investigated, particularly for $Li_2O_2$.[22–25] For example, dopants such as Ba, $Na^+$, and $K^+$ have been reported to promote lithium vacancy formation and lower the charging overpotential.[26–31] Meanwhile, electrolyte additives have been used to modify the structural characteristics of $Li_2O_2$; fluorinated amide-based electrolytes have been experimentally shown to induce the formation of lithium-deficient $Li_{2-x}O_2$ and maintain a low charging voltage.[32] Also, the addition of chloride ions to the electrolyte has been reported to form chloride-containing $Li_2O_2$ deposits, leading to a significant improvement in the energy capacity.[33] These results suggest that chemical modification through electrolyte composition and doping is a promising strategy for tailoring the structure and stability of discharge products, thereby improving the energy efficiency and cyclability of LOBs.

However, most of these studies have focused on $Li_2O_2$. In the case of $Li_2CO_3$, while its charge transport properties have been investigated,[6–9,34–38] the intrinsic defect energetics of $Li_2CO_3$ have not been systematically characterized. Furthermore, the use of dopants to control the thermodynamic stability of $Li_2CO_3$ remains thoroughly unexplored. This motivates the identification of dopants that can modify $Li_2CO_3$ defect energetics under battery-relevant conditions. Among the various dopant candidates, fluorine is particularly noteworthy for several reasons. First, from a practical standpoint, fluorinated electrolytes may serve as a source of fluorine-containing species under realistic battery operating conditions. Although the effects of fluorinated electrolytes on $Li_2CO_3$ have not been directly examined, their reported role in promoting the formation of lithium-deficient $Li_{2-x}O_2$ discharge products provides practical motivation for investigating fluorine as a dopant candidate.[32] In this respect, fluorine differs from many other dopants that would require deliberate synthetic incorporation. Second, owing to its similar ionic radius to oxygen and its exceptional electronegativity, fluorine substitution at oxygen sites is expected to selectively perturb the local bonding environment around lithium ions, potentially modifying cation sublattice stability without wholesale structural disruption. Third, while halogen doping effects including fluorine have been theoretically explored in $Li_2O_2$,[22,23] a systematic investigation of fluorine doping in $Li_2CO_3$ remains absent. Extending such fluorine doping strategies from $Li_2O_2$ to $Li_2CO_3$ may therefore provide a promising route for controlling the stability of $Li_2CO_3$ as a discharge product phase.

In this study, first-principles calculations were performed to systematically analyze the formation energies of intrinsic point defects in $Li_2CO_3$ and to assess the thermodynamic role of fluorine doping in modifying the vacancy formation energies of both $Li_2CO_3$ and $Li_2O_2$. The atomic-scale mechanism by which fluorine is incorporated into these discharge products under realistic battery operating conditions remains unresolved; the fluorine substitution energies calculated here are referenced to LiF as the thermodynamically stable fluorine reservoir and therefore represent upper-bound estimates of the incorporation cost. Motivated by experimental observations that fluorine-containing electrolyte species can nonetheless be found in association with discharge products, this study assumes fluorine solid–solution incorporation as a working hypothesis and evaluates the resulting changes in defect formation energies, [32,33] rather than asserting the incorporation mechanism itself.

Through this approach, we clarify the defect thermodynamics underlying the stability of $Li_2CO_3$ and examine how fluorine doping affects the stability of these discharge products. This work offers thermodynamic insight into how fluorine doping modifies the defect energetics of $Li_2CO_3$ and $Li_2O_2$, which may inform doping design strategies for controlling the stability of lithium-based discharge products.

## METHODS

### DFT calculations and structure formation

All calculations were performed within the framework of density functional theory (DFT) as implemented in the Vienna Ab initio Simulation Package (VASP).[39–41] The electron–ion interactions were described by the projector augmented-wave (PAW) method.[42] For structural relaxations, the exchange-correlation energy was treated using the Perdew–Burke–Ernzerhof (PBE) functional within the generalized gradient approximation (GGA).[43] For Li, the Li_sv PAW potential including both 1s and 2s states in the valence was employed, whereas default PAW datasets were used for O, C, and F. The plane-wave cutoff energy was set to 520 eV, and Brillouin zone integrations were performed using Γ-centered k-point grids generated with a reciprocal-space resolution of 0.025 $Å^{-1}$. The convergence criteria were defined as $10^{-7}$ eV for the electronic self-consistent field step and 0.005 eV/Å for the residual ionic forces. Spin polarization was included in all calculations, as its role becomes evident upon fluorine substitution in $Li_2CO_3$. The fluorine-substituted supercell exhibits a non-negligible total magnetization of approximately 0.65 $\mu_B$, indicating that fluorine substitution induces local unpaired electron density through charge compensation effects. The corresponding spin density is localized predominantly on the carbon atom adjacent to the substituted fluorine atom and is partially distributed over the remaining oxygen atoms within the same carbonate-derived unit. This finding supports the use of spin-polarized calculations for accurately describing the electronic structure of fluorine-doped $Li_2CO_3$. Upon introduction of a lithium vacancy, the total magnetization decreases to nearly zero, consistent with electronic compensation accompanying lithium removal. In contrast, carbon vacancy formation leads to a slight increase in magnetization, and oxygen vacancy formation results in only a minor change in magnetization, corroborating the selective nature of fluorine doping toward cation vacancy formation rather than anion vacancy formation. These trends in magnetization are qualitatively consistent with the computed vacancy formation energies and provide electronic evidence for the selective modulation of defect energetics by fluorine substitution. Gaussian smearing with a width of 0.05 eV was applied for self-consistent electronic calculations. For the bulk unit cells, both lattice parameters and internal atomic coordinates were fully relaxed.

For the monoclinic unit cell of $Li_2CO_3$, the optimized lattice parameters were a = 8.38 Å, b = 5.02 Å, c = 6.30 Å, β = 114.15°, which are in good agreement with crystallographic data (a = 8.36 Å, b = 4.97 Å, c = 6.19 Å, β = 114.79°).[44] For the hexagonal unit cell of $Li_2O_2$, the optimized lattice parameters (a = b = 3.16 Å, c = 7.68 Å) were consistent with reported experimental values (a = b = 3.14 Å, c = 7.65 Å).[45] Based on the optimized bulk unit cells, a 2 × 2 × 2 supercell (192 atoms) for $Li_2CO_3$ and a 4 × 4 × 2 supercell (256 atoms) for $Li_2O_2$ were generated.

Intrinsic defect energetics of pristine $Li_2CO_3$ were evaluated by considering both neutral and charged defects. Fluorine substitution was introduced separately at the two inequivalent oxygen sites in the carbonate unit of $Li_2CO_3$, corresponding to a nominal fluorine concentration of 1.04% on the oxygen sublattice. For fluorine-substituted $Li_2CO_3$, both neutral and charged vacancy configurations were considered to evaluate the effect of fluorine doping on defect energetics over different Fermi level conditions. In $Li_2O_2$, fluorine substitution was treated as a neutral defect and modeled by replacing one peroxide dimer, corresponding to a nominal substitution concentration of 1.56% on the peroxide-dimer sublattice, as this configuration is energetically more favorable than alternative substitutions.[22,23] For $Li_2O_2$, only neutral vacancy configurations were considered to compare the voltage-dependent vacancy formation energies between pristine and fluorine-substituted structures.

In the fluorine-doped structures, vacancy sites were selected in the vicinity of the substituted fluorine atom to evaluate the local influence of fluorine substitution on vacancy formation energies. Complex vacancy defects, such as $V^0_{Li_2CO_3}$, were evaluated using direct supercell calculations by removing one $Li_2CO_3$ formula unit from the lattice as a clustered vacancy. This direct approach explicitly includes local structural relaxation and interactions among the constituent vacancies, instead of approximating the complex defect energy by summing the formation energies of individual charged vacancies.

The supercells were optimized at the PBE level with fixed lattice parameters, allowing only internal atomic coordinates to relax. The PBE-relaxed structures were used for single-point total energy calculations using the Heyd–Scuseria–Ernzerhof (HSE) functional to evaluate defect formation energies.[46–48] The HSE mixing parameter was set to $\alpha = 0.35$, following previous work,[49] in which this value was adopted as an intermediate choice within the range of reported GW band gap estimates for $Li_2O_2$. Γ-centered 2 × 2 × 2 k-point meshes were employed for the supercells, and the electronic convergence criterion was set to $10^{-6}$ eV.

**Defect formation energy**

The defect formation energy of a defect $X$ in charge state $q$ was defined as follows:[50]

$$E^f[X^q] = E_{\text{tot}}[X^q] - E_{\text{tot}}[\text{ref}] + \sum_i n_i\mu_i + qE_{\text{F}} + E_{\text{corr}} \quad (1)$$

Where $E_{\text{tot}}[X^q]$ and $E_{\text{tot}}[\text{ref}]$ denote the total energies of the defective and reference supercells, respectively. The reference supercell is the pristine supercell for intrinsic defects and the fluorine-substituted supercell for vacancy defects in doped systems. The term $n_i$ represents the number of atoms of species $i$ added to or removed from the reference structures; $n_i > 0$ when atoms are removed and $n_i < 0$ when atoms are added. $\mu_i$ is the chemical potential of species $i$.

The lithium chemical potential was defined to account for the electrochemical conditions of LOBs environment. In this environment, the applied cell voltage $U$ exists between the lithium metal anode and the discharge products formed at the cathode. Accordingly, the lithium chemical potential depends on $U$ and was expressed as:[51]

$$\mu_{\mathrm{Li}}(U) = \mu_{\mathrm{Li}}^{\mathrm{bulk}} - \mathrm{e}U \quad (2)$$

where $\mu_{\mathrm{Li}}^{\mathrm{bulk}}$ is the energy per atom of bulk lithium metal at $U = 0$, and e is the elementary charge. This voltage-dependent definition of $\mu_{Li}$ was applied to both discharge products. The equilibrium voltages of $Li_2O_2$ and $Li_2CO_3$ were used as reference electrochemical conditions for evaluating vacancy formation energies. These voltages correspond to the following cathode reactions:[6,15–18]

$$2\mathrm{Li}^+ + 2\mathrm{e}^- + \mathrm{O}_2 \rightleftharpoons \mathrm{Li}_2\mathrm{O}_2, \qquad U_{\mathrm{eq}} = 2.96\ \mathrm{V\ vs\ Li}^+/\mathrm{Li} \quad (3)$$

$$4\mathrm{Li}^+ + 4\mathrm{e}^- + 2\mathrm{CO}_2 + \mathrm{O}_2 \rightleftharpoons 2\mathrm{Li}_2\mathrm{CO}_3, \qquad U_{\mathrm{eq}} = 3.82\ \mathrm{V\ vs\ Li}^+/\mathrm{Li} \quad (4)$$

The oxygen chemical potential was fixed by equilibrium with molecular $O_2$ in the atmosphere, with a correction applied to account for the well-known overbinding error of gas-phase $O_2$ in DFT calculations; the correction was derived from the experimental formation enthalpy of $Li_2O_2$ as described in **Supporting Information S1**. This correction was used to define the oxygen chemical potential under the $Li_2O_2$-dominated electrochemical conditions assumed here, rather than to correct the covalent C–O bonding in $Li_2CO_3$. The carbon chemical potential was determined by the thermodynamic stability condition of $Li_2CO_3$. The fluorine chemical potential was primarily defined using LiF as the thermodynamic reference reservoir, assuming equilibrium among $Li_2O_2$, $O_2$, and LiF:[23]

$$\mu_{\mathrm{F}} = E(\mathrm{LiF}) - \frac{1}{2}E(\mathrm{Li}_2\mathrm{O}_2) + \frac{1}{2}E(\mathrm{O}_2) \quad (5)$$

This represents a conservative upper-bound estimate of the substitution cost under equilibrium bulk conditions; additional fluorine chemical potentials were calculated using $F_2$ and $OF_2$ as fluorine-rich reference states. The detailed chemical potential calculations, phase stability boundaries, and definitions of the alternative fluorine reference states are described in **Supporting Information S1**.

The Fermi level $E_{\mathrm{F}}$ is referenced to the valence band maximum (VBM) of the bulk system and was varied from 0 to the calculated band gap. $E_{\mathrm{corr}}$ denotes the finite-size electrostatic correction term for charged-defect supercell calculations. Finite-size electrostatic corrections were applied to charged defects using the Freysoldt–Neugebauer–Van de Walle (FNV) scheme, as implemented in the pydefect package. [52–54]

## RESULTS and DISCUSSION

### Structural Properties and intrinsic defect energetics of $Li_2CO_3$

The structural model used in this study is a monoclinic $Li_2CO_3$ supercell with space group C2/c, as presented in **Figure 1(a,b)** together with the local structure of the carbonate unit and the distinct atomic sites considered for defect modeling. The Wyckoff positions and fractional coordinates of all inequivalent atomic sites in the optimized structure are summarized in **Table S2**. $Li_2CO_3$ consists of ionic interactions between $\mathrm{Li}^{+}$ cations and $\mathrm{CO}_3^{2-}$ anions. Within each $\mathrm{CO}_3^{2-}$ unit, the carbon atom is coordinated by three oxygen atoms in a trigonal planar geometry, reflecting the covalent character of the C–O bonds.

The oxygen atoms in the carbonate unit can be classified into two inequivalent sites, denoted as O(f) and O(e). According to the optimized structure, the C–O bond length at the O(f) site is 1.31 Å and is coordinated by three neighboring lithium ions. In contrast, the C–O bond length at the O(e) site is 1.29 Å and is coordinated by two neighboring lithium sites. Due to its higher coordination with lithium ions, the O(f) site exhibits a relatively lower C–O bond order, resulting in an elongated bond length compared with O(e). The optimized structure of $Li_2O_2$ is provided in **Figure S2(a)**.

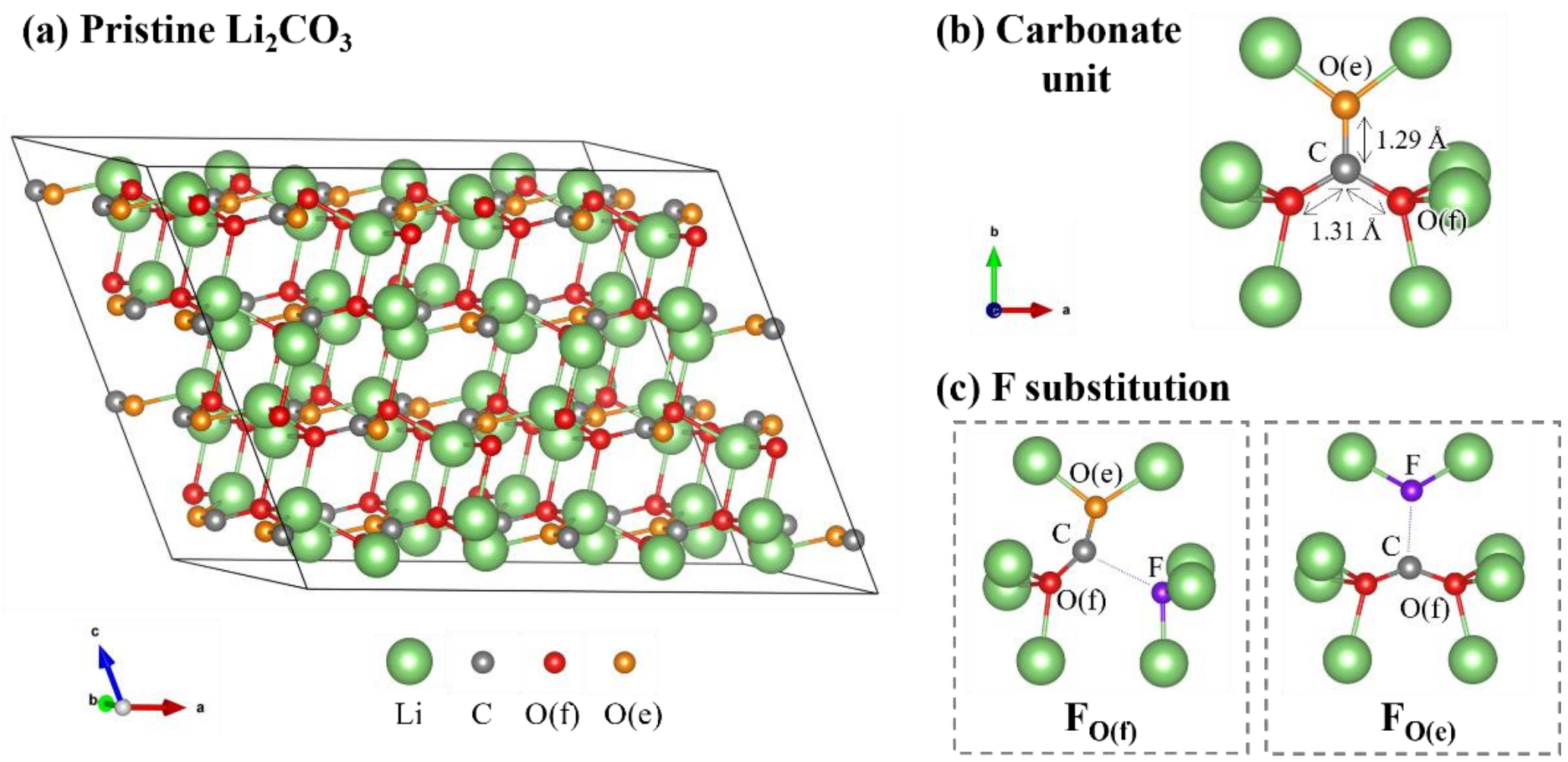

**Figure 1**. (a) Crystal structure of the optimized 2 × 2 × 2 $Li_2CO_3$ supercell with two inequivalent O sites. (b) Carbonate unit showing the inequivalent O(f) and O(e) sites. (c) optimized local structures of F-substituted at O(f) and O(e) sites. Dashed lines indicate the original C–O bonds before F-substitution.

The formation energies of intrinsic point defects in bulk $Li_2CO_3$ were calculated as a function of the Fermi level across its band gap (~7.98 eV) as shown in **Figure 2**. The defects considered include vacancies and interstitials as well as complex vacancy defects in which multiple atoms are removed together as a single entity. To examine the dependence of defect energetics on the lithium chemical potential, three representative voltage conditions were defined: 0 V, 2.96 V and 3.82 V vs. $\mathrm{Li^+/Li}$. Although practical charging can occur at higher voltages due to overpotential, the chemical potentials in this study were defined using equilibrium voltages. While the main discussion focuses on voltage-dependent lithium chemical potentials, additional defect formation energies based on the phase diagram of the Li–C–O system are provided in **Figure S3** as a complementary reference.

**Figure 2**(a) shows the defect formation energies under Li-rich conditions at 0 V, where the lithium chemical potential is referenced to equilibrium with lithium metal at the anode. Under this condition, $\mathrm{Li_i^+}$ , $\mathrm{V_{CO_2}^0}$ and $\mathrm{V_{CO_3}^{2+}}$ exhibit relatively low formation energies. The low formation energy of $\mathrm{Li_i^+}$ is consistent with previous study,[3] which reported $\mathrm{Li_i^+}$ as the most stable lithium-related defect species under low voltage conditions. In addition, the calculated formation energy of $\mathrm{V_{Li}^0}$ was approximately 5 eV, which agrees with the previously reported value.[6] These comparisons support the reliability of these results. The relatively low formation energies of $\mathrm{V_{CO_2}^0}$ and $\mathrm{V_{CO_3}^{2+}}$ indicate that carbonate unit becomes thermodynamically unstable under Li-rich conditions. However, this trend should be interpreted as a thermodynamic indicator based on bulk defect formation energies and does not directly define the actual decomposition pathway.

**Figure 2**(b) shows the defect formation energies at 2.96 V, corresponding to the equilibrium voltage for $Li_2O_2$. Under this condition, the defect behavior was mainly dominated by $\mathrm{Li_i^+}$ and $\mathrm{V_{Li}^-}$, whereas carbon- and oxygen-related intrinsic defects remained thermodynamically unfavorable.

At 3.82 V, corresponding to the theoretical equilibrium voltage of $Li_2CO_3$, **Figure 2**(c) shows that $\mathrm{V_{Li}^-}$ becomes the dominant defect species. In contrast, carbon- and oxygen-related defects retained high formation energies, indicating that they are energetically unfavorable near the $Li_2CO_3$ decomposition voltage.

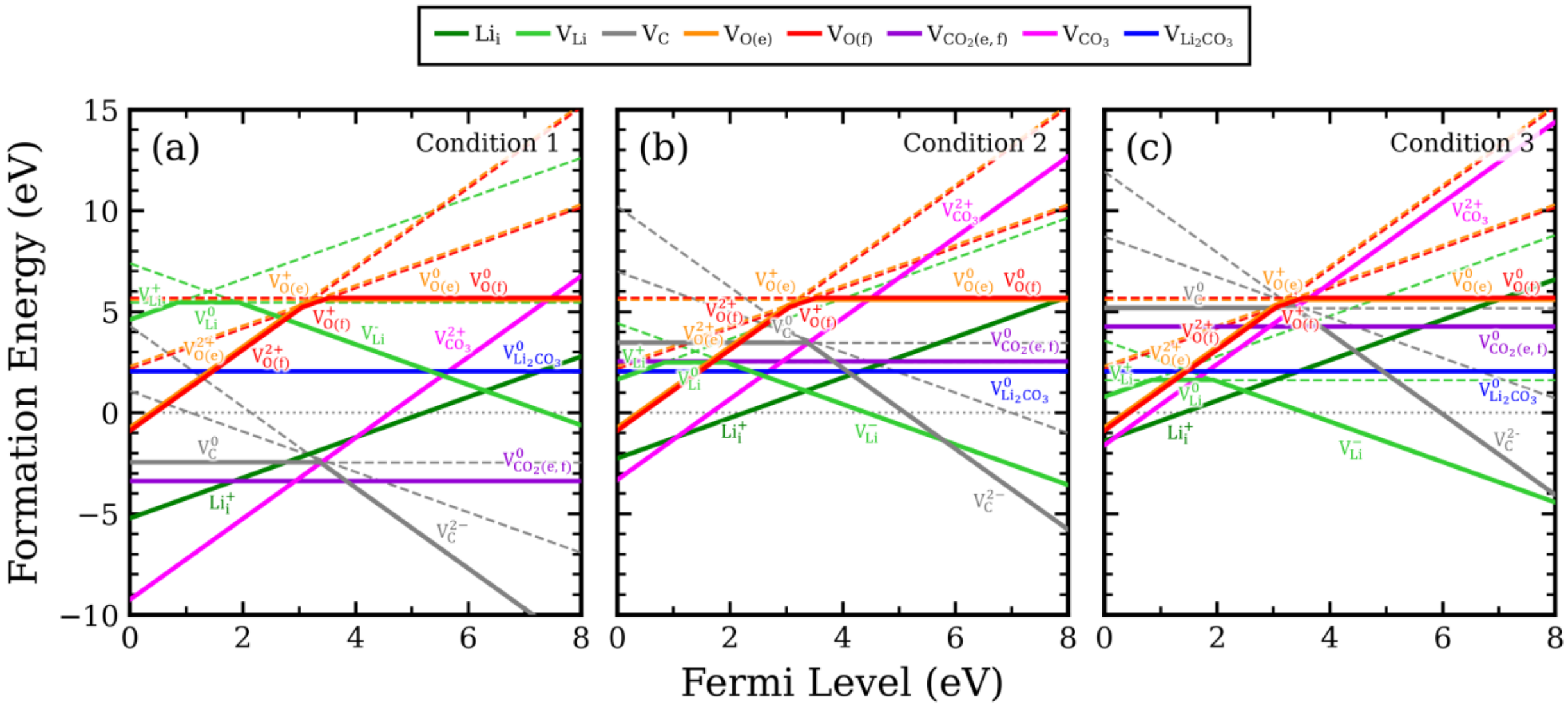

**Figure 2.** Formation energies of intrinsic point defects in bulk $Li_2CO_3$ as a function of the Fermi level under voltage-dependent Li chemical potential: (a) 0 V, (b) 2.96 V, (c) 3.82 V vs. $Li^+/Li$. The Fermi level is referenced to the valence band maximum and varied across the calculated band gap of $Li_2CO_3$. The thermodynamically most stable charge states are represented by solid lines, with higher energy states shown as dashed lines.

The voltage-dependent defect behavior of $Li_2CO_3$ was generally governed by lithium-related defects, while carbon- and oxygen-related defects were relatively difficult to form. In particular, the dominant formation of $V_{Li}^-$ near and above the equilibrium voltage of $Li_2O_2$ suggests that $Li^+$ extraction accompanied by electronic compensation may be thermodynamically favored during charging. This supports that lithium removal has a relatively low thermodynamic cost, consistent with previously proposed oxidative process during the initial decomposition of $Li_2CO_3$.

For carbon monoxide defects, reliable bulk formation energies could not be determined in this study, presumably because these configurations do not correspond to stable point defects in crystalline $Li_2CO_3$. This is consistent with experimental observations, which suggest that carbon monoxide does not directly evolve from $Li_2CO_3$ during electrochemical oxidation.[8] Rather carbon monoxide formation has been attributed to secondary reactions involving reactive intermediates, such as singlet oxygen generated during $Li_2CO_3$ decomposition, which subsequently attack the carbon substrate or electrolyte.

The results in this section demonstrate a quantitative energy hierarchy suggesting that lithium removal through vacancy formation within the rigid carbonate lattice has a relatively low thermodynamic cost under charging conditions, thereby providing a thermodynamic basis for its proposed involvement in the initiation of $Li_2CO_3$ decomposition. This intrinsic defect analysis provides a thermodynamic basis for the following section, which discusses how the doping strategy affects the defect energetics of $Li_2CO_3$.

**Effect of fluorine doping on defect formation energetics in discharge products**

To evaluate the thermodynamic effect of fluorine doping on vacancy formation in discharge products, we first calculated the substitution energetics of fluorine in $Li_2CO_3$ and $Li_2O_2$. For a consistent comparison between the two discharge product phases, the same fluorine chemical potential was applied to both systems. Using LiF as the fluorine reference reservoir, the fluorine substitution energies were calculated to be 4.75 eV and 5.01 eV for O(f) and O(e) sites in $Li_2CO_3$ and 2.10 eV for $Li_2O_2$. The energetic cost of fluorine substitution, however, is sensitive to the chosen fluorine reservoir. Under fluorine-rich reference conditions using $F_2$ and $OF_2$, the substitution energies decreased substantially to 1.76–1.81 eV for the O(f) site and 2.02–2.08 eV for the O(e) site in $Li_2CO_3$, while the corresponding values for $Li_2O_2$ become negative, ranging from −0.89 to −0.84 eV (**Table S3**). The LiF-based values therefore represent conservative upper-bound estimates under equilibrium bulk conditions. On this basis, while the LiF-based values represent upper-bound estimates, fluorine-rich electrochemical environments may yield lower effective substitution energies owing to increased local fluorine activity at the electrolyte–discharge product interface. Accordingly, this study assumes the presence of pre-incorporated fluorine dopants and focuses on the relative changes in defect formation behavior between pristine and fluorine-doped systems. This approach is in line with previous theoretical studies that evaluated dopant effects in $Li_2O_2$ under the assumption of pre-incorporated species.[27,28] It also provides meaningful thermodynamic insight into how fluorine modifies the defect landscape of discharge products regardless of the precise incorporation pathway.

**Figure 3** shows the formation energies of vacancies in pristine and fluorine-doped $Li_2CO_3$ as a function of the Fermi level, using chemical potentials fixed at the equilibrium voltage of $Li_2CO_3$ (3.82 V). The voltage was fixed to isolate the Fermi-level dependence of the defect charge states. The separate calculated formation energies of charged vacancies for fluorine substitution at the O(f) and O(e) sites are summarized in **Figure S4**. **Figure 1(c)** shows the optimized local geometries around fluorine substitution of $Li_2CO_3$ at the inequivalent oxygen sites. The local structural and electronic changes induced by fluorine doping are summarized in **Supporting Information S4**. As $Li_2CO_3$ requires higher overpotentials for decomposition, its decomposition has been reported to involve an electrochemical $Li^+$ extraction process prior to the chemical dissociation of carbonate ions. [6,9] Consequently, the energetics of lithium removal and the structural integrity of the carbonate framework are relevant to understanding the thermodynamic characteristics of $Li_2CO_3$.

In the fluorine-doped structure, the lithium vacancies formation energies decreased in the intermediate Fermi-level region, whereas the difference between the pristine and fluorine-doped structures remained relatively small near the VBM and CBM. This indicates that the tendency for lithium vacancy formation depends on the Fermi level. The carbon vacancies were energetically more favorable in the fluorine-doped structure over the entire Fermi-level range. For oxygen vacancies, only minor differences were observed, and fluorine substitution even made oxygen vacancy formation less favorable in the low Fermi-level region.

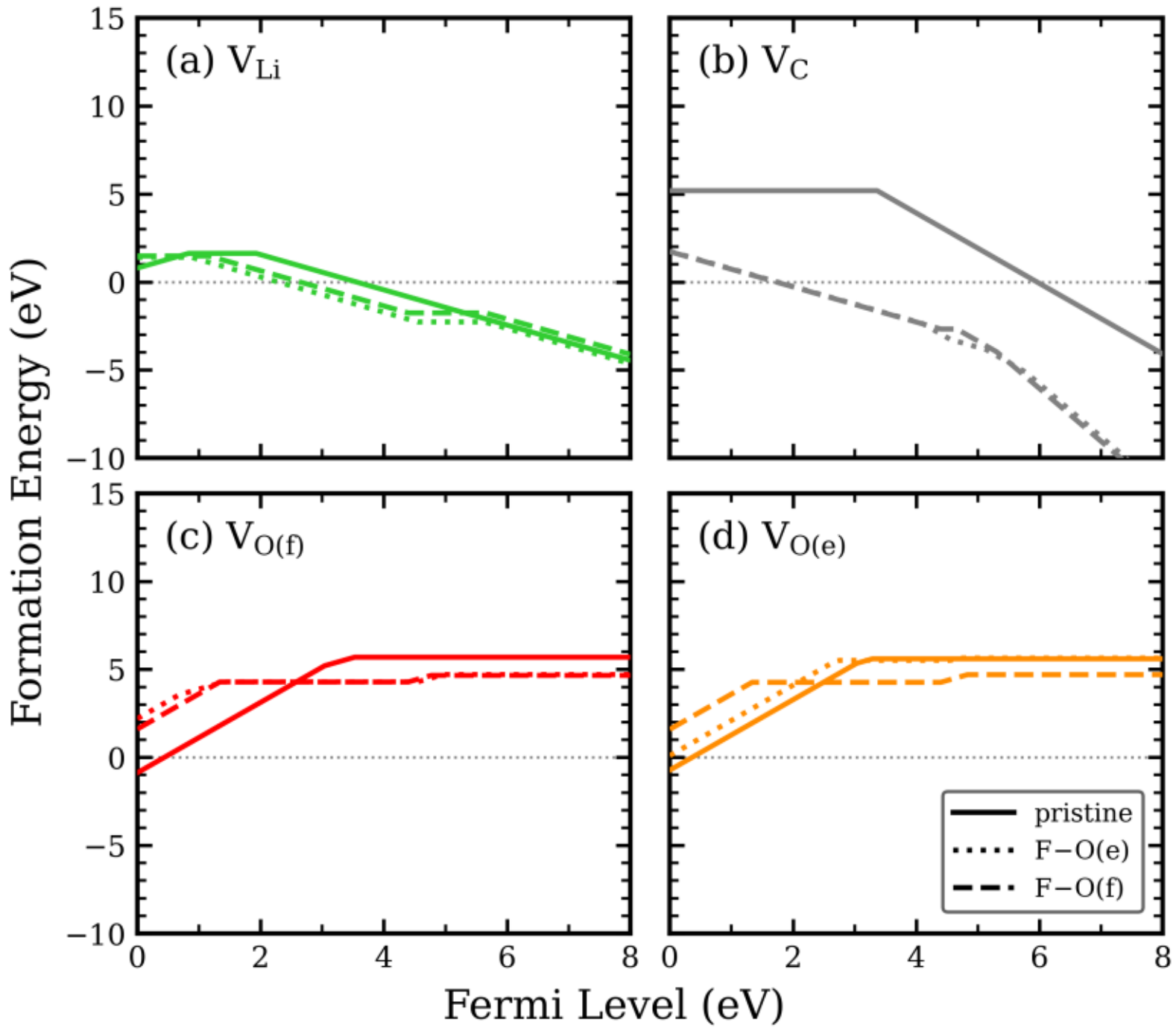

**Figure 3.** Formation energies of vacancies in pristine (solid line) and F-doped $Li_2CO_3$ (dashed and dotted lines) as a function of applied voltage, 3.82 V: (a) Li vacancy, (b) C vacancy, (c) O vacancy at O(f) site and (d) O vacancy at O(e) site. Only the lowest energy charge states are shown for each defect.

To quantitatively evaluate the fluorine doping effect, the defect energetics were compared using neutral vacancy configurations. Upon fluorine doping, the lithium vacancy formation energy decreased from 1.63 eV to −1.75 eV ($\Delta E \approx -3.38$ eV) for fluorine substitution at the O(f) site and to −2.27 eV ($\Delta E \approx -3.90$ eV) at the O(e) site, indicating a reduced thermodynamic cost for lithium removal. Simultaneously, the carbon vacancy formation energy dropped from 5.18 eV to −2.67 eV ($\Delta E \approx -7.85$ eV) and to −2.80 eV ($\Delta E \approx -7.98$ eV) for fluorine substitution at the O(f) and O(e) sites, respectively. Given that the complete decomposition of $Li_2CO_3$ is limited by the dissociation of the rigid carbonate network,[6,36] this decrease suggests a partial weakening of the framework's thermodynamic stability. In contrast, the oxygen vacancy formation remains unlikely to serve as a dominant bulk defect pathway. Overall, fluorine doping lowers the formation energy of neutral lithium vacancies, although the stability of charged lithium vacancies remains dependent on the Fermi level, while the formation energy of carbon vacancies is consistently reduced over the entire Fermi-level range. These results indicate that fluorine doping not only partially facilitates lithium removal under certain electronic conditions but also weakens the thermodynamic stability of the carbon framework.

Subsequently, **Figure 4** shows the formation energies of neutral vacancies in $Li_2O_2$ as a function of applied voltage. The voltage range was set to 2.0–4.5 V (vs. $Li^+/Li$) to examine the voltage dependence of the vacancy energetics under practically relevant operating conditions, including the equilibrium voltage of $Li_2O_2$ and higher-voltage region associated with charging overpotentials. The local structure around the fluorine substitution is shown in **Figure S2**(b). Fluorine doping reduced the lithium vacancy formation energies compared with the pristine state, whereas the oxygen vacancy formation energy remained largely unchanged.

To assess this effect, the vacancy formation energies were compared at the equilibrium voltage of $Li_2O_2$, 2.96 V. Under this condition, fluorine doping reduced the lithium vacancy formation energy from 1.02 eV to −0.66 eV ($\Delta E \approx -1.68$ eV) and from 0.89 eV to −0.54 eV ($\Delta E \approx -1.43$ eV) for the two inequivalent lithium sites. Among the proposed pathways for $Li_2O_2$ decomposition during charging, one possible route involves $Li^+$ extraction and the formation of lithium-deficient $Li_{2-x}O_2$ phases, where lithium vacancies are key defects influencing both thermodynamic stability and charge-transport properties.[18,19,55–57] Therefore, the reduction in lithium vacancy formation energy observed here suggests that fluorine doping thermodynamically favors lithium removal, a step involved in the formation of lithium-deficient $Li_{2-x}O_2$ phases during charging. This interpretation is qualitatively consistent with experimental observations that fluorine-containing electrolytes promote the formation of poorly crystalline, lithium-deficient $Li_{2-x}O_2$ discharge products that can be decomposed at lower charging voltages.[32] These results are consistent with the possibility that fluorine doping may contribute to reducing the thermodynamic cost associated with the formation of lithium-deficient $Li_{2-x}O_2$ phases.

In contrast, the oxygen vacancy formation energy changed only marginally by 0.09 eV after fluorine doping and remained relatively high. This indicates that the peroxide sublattice in $Li_2O_2$ maintains its structural integrity, confirming that the dopant effect acts selectively on the lithium sublattice rather than inducing a uniform destabilization of the entire crystal lattice.

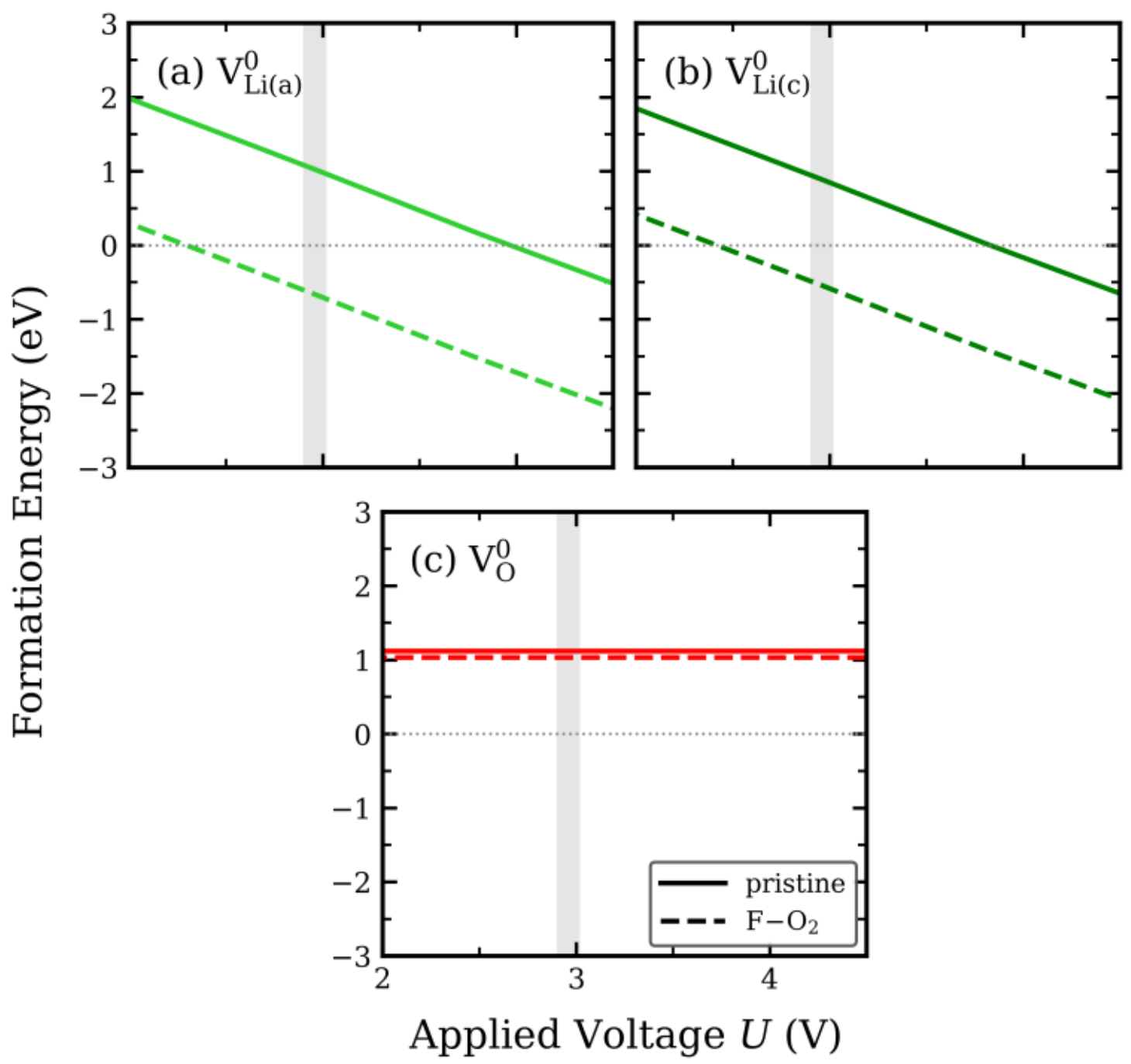

**Figure 4.** Formation energies of neutral vacancies in pristine and fluorine-doped $Li_2O_2$ as a function of applied voltage $U$: (a) Li vacancy at Li(a) site, (b) Li vacancy at Li(c) site, and (c) O vacancy. Li(a) and Li(c) denote two inequivalent Li sites, as defined in **Figure S2**. The gray shaded region indicates the equilibrium voltage of $Li_2O_2$.

Comparing the results for $Li_2CO_3$ and $Li_2O_2$ reveals a consistent trend: fluorine doping selectively reduces the lithium vacancy formation energies in both systems while exerting limited influence on oxygen vacancies. This demonstrates that fluorine at oxygen sites has a stronger effect on cation sublattices rather than the anion sublattices. This selectivity can be attributed to the strong electronegativity of fluorine, which perturbs the local electrostatic environment around lithium sites upon substitution at oxygen sites. These findings imply that the experimentally reported influence of fluorinated electrolytes on $Li_2O_2$ discharge products extends to $Li_2CO_3$ by reducing the thermodynamic stability of its bulk structure and facilitating vacancy formation.[32]

Notably, the magnitude of the fluorine doping effect differed markedly between the two systems. In $Li_2CO_3$, the reduction in neutral lithium vacancy formation energy ($\Delta E \approx -3.4$ eV to $-3.9$ eV) was approximately twice the reduction observed in $Li_2O_2$ ($\Delta E \approx -1.4$ eV to $-1.7$ eV). This difference likely reflects the stronger perturbation induced by fluorine substitution within the rigid carbonate framework, where the highly covalent C–O bonds concentrate the electrostatic effect of fluorine more locally around the neighboring lithium sites. Furthermore, $Li_2CO_3$ exhibits an additional fluorine doping effect that has no counterpart in $Li_2O_2$: a dramatic reduction in neutral carbon vacancy formation energy ($\Delta E \approx -7.9$ eV). Given that $Li_2CO_3$ decomposition is limited by dissociation of the carbonate framework, this finding suggests that fluorine doping not only facilitates lithium removal but may also partially destabilize the carbonate network itself, potentially providing an additional pathway that is absent in pristine $Li_2CO_3$.[6,36]

These findings indicate that fluorine doping modifies the defect energetics of $Li_2CO_3$ through a richer set of mechanisms than in $Li_2O_2$, including a reduction in carbon vacancy formation energy that is absent in $Li_2O_2$. These results suggest that the defect-energetic effects of fluorine, previously studied in $Li_2O_2$, may extend to $Li_2CO_3$, potentially offering an additional avenue for modifying the thermodynamic stability of parasitic discharge products in LOBs.

## CONCLUSIONS

In this study, first-principles calculations revealed that lithium-related defects are the dominant intrinsic defects in $Li_2CO_3$, while carbon- and oxygen-related vacancy formation is energetically unfavorable near the $Li_2CO_3$ equilibrium voltage. These results suggest that lithium removal is thermodynamically more favorable than direct carbonate framework dissociation through bulk point defects. This intrinsic resistance of pristine $Li_2CO_3$ to bulk defect formation indicates the need to modify its defect thermodynamics. The present results show that fluorine doping in $Li_2CO_3$ selectively decreases the lithium vacancy formation energies under certain conditions, whereas carbon vacancy formation energies are reduced across the entire range examined. The oxygen vacancy formation is only weakly affected. This cation-selective destabilization, attributable to the strong electronegativity of fluorine perturbing the local electrostatic environment around lithium sites, not only lowers the thermodynamic cost of lithium removal but also partially weakens the carbonate framework itself, an effect with no direct counterpart in $Li_2O_2$. Notably, the magnitude of the fluorine doping effect on lithium vacancy formation was approximately twice as large in $Li_2CO_3$ ($\Delta E \approx -3.4$ eV to $-3.9$ eV) as in $Li_2O_2$ ($\Delta E \approx -1.4$ eV to $-1.7$ eV), suggesting that the rigid carbonate framework amplifies the local electrostatic perturbation induced by fluorine substitution. Together with the results for $Li_2O_2$, these findings suggest that fluorine doping modulates the thermodynamic stability of both discharge products through a common mechanism of cation sublattice destabilization, while additionally destabilizing the carbonate framework in $Li_2CO_3$. These results are consistent with the possibility that fluorine doping strategies explored for $Li_2O_2$ may also be applicable to $Li_2CO_3$. Future work incorporating charge-state-dependent defect behavior, polaron dynamics, and kinetic barriers will be essential to fully elucidate the decomposition mechanism under realistic electrochemical conditions and to translate these thermodynamic insights into practical battery design.

## ASSOCIATED CONTENT

### Supporting Information

The supporting information is available free of charge on the ACS Publications website.

Chemical potential calculations, structural information, additional defect formation energy results, structural and electronic changes in $Li_2CO_3$ (PDF)

## AUTHOR INFORMATION

Corresponding Author

Teruyasu Mizoguchi — *Institute of Industrial Science, The University of Tokyo, Tokyo 153-8505, Japan*, and *Department of Materials Engineering, The University of Tokyo, Tokyo 113-8656, Japan*; orcid.org/0000-0003-3712-7307; Email: teru@iis.u-tokyo.ac.jp

Authors

Youjeong Choi — *Department of Materials Engineering, The University of Tokyo, Tokyo 113-8656, Japan*

Tasuku Sugiura — *Institute of Industrial Science, The University of Tokyo, Tokyo 153-8505, Japan*

Keisuke Mukai — *National Institute for Fusion Science, Gifu 509-5292, Japan*

Nanako Ishihara — *Research Center for Solar Energy Chemistry, Graduate School of Engineering Science, The University of Osaka, Osaka 560-8531, Japan*

Shuji Nakanishi — *Research Center for Solar Energy Chemistry, Graduate School of Engineering Science, The University of Osaka, Osaka 560-8531, Japan*, and *Innovative Catalysis Science*

*Division, Institute for Open Transdisciplinary Research Initiatives, The University of Osaka, Osaka 565-0871, Japan*

Author Contributions

Y. C. performed the first-principles calculations, analyzed the data, prepared the figures, and wrote the original draft. T. S, K. M, N. I, S. N. contributed to data interpretation and manuscript review. T. M. supervised the research, contributed to the interpretation of the results, and revised the manuscript. All authors have approved the final version of the manuscript.

Funding Sources

This work was supported by the Japan Science and Technology Agency (JPMJGX24S0).

Notes

The authors declare no competing financial interest.

## ACKNOWLEDGMENT

This work was supported by the Japan Science and Technology Agency (JST) as part of Green Technologies of Excellence (GteX), Grant Number JPMJGX24S0. The computation was carried out using the computer resource offered by Research Institute for Information Technology, Kyushu University.

**Supporting Information**

# Intrinsic Defect Energetics and Fluorine Doping Effects in $Li_2CO_3$ and $Li_2O_2$: A First-Principles Study

*Youjeong Choi*[1], *Tasuku Sugiura*[2], *Keisuke Mukai*[3], *Nanako Ishihara*[4], *Shuji Nakanishi*[4, 5], *and Teruyasu Mizoguchi*[1, 2, *]

[1] Department of Materials Engineering, The University of Tokyo, Tokyo 113-8656, Japan

[2] Institute of Industrial Science, The University of Tokyo, Tokyo 153-8505, Japan

[3] National Institute for Fusion Science, Gifu 509-5292, Japan

[4] Research Center for Solar Energy Chemistry, Graduate School of Engineering Science, The University of Osaka, Osaka 560-8531, Japan

[5] Innovative Catalysis Science Division, Institute for Open Transdisciplinary Research Initiatives, The University of Osaka, Osaka 565-0871, Japan

*Corresponding author. E-mail: teru@iis.u-tokyo.ac.jp

## S1. The calculation of chemical potential

For the voltage-dependent analysis, the lithium chemical potential was defined using **Eq. (2)** in the main text. The same voltage-dependent $\mu_{Li}(U)$ was applied to both $Li_2CO_3$ and $Li_2O_2$.

The oxygen chemical potential was corrected to account for the well-known overbinding error of gas-phase $O_2$ in DFT calculations. Although the main target phase of this study is $Li_2CO_3$, the $O_2$ correction was derived from $Li_2O_2$ following previous studies,[1,2] and the same corrected $O_2$ energy was consistently used for both $Li_2CO_3$ and $Li_2O_2$. Following previous studies, the correction was determined from the difference between the calculated and experimental formation enthalpies of $Li_2O_2$. The experimental value was taken as $\Delta H_f^{\mathrm{exp}}(\mathrm{Li_2O_2}, 300\ \mathrm{K}) = -6.57$ eV.[3] In this correction scheme, the DFT total-energy difference was used as an approximation to the formation enthalpy, while finite-temperature free-energy contributions of the solid phases were not explicitly considered. The $O_2$ correction procedure was expressed as follows:

$$\Delta H_f^{\mathrm{cal}}(\mathrm{Li_2O_2}) = E^{\mathrm{DFT}}(\mathrm{Li_2O_2}) - 2\,E^{\mathrm{DFT}}(\mathrm{Li_{bulk}}) - E^{\mathrm{DFT}}\,(\mathrm{O_2})$$

$$\Delta E(\mathrm{O_2}) = \Delta H_f^{\mathrm{cal}}(\mathrm{Li_2O_2}) - \Delta H_f^{\mathrm{exp}}(\mathrm{Li_2O_2}, 300\ \mathrm{K})$$

$$E^{\mathrm{corr}}\,(\mathrm{O_2}) = E^{\mathrm{DFT}}(\mathrm{O_2}) + \Delta E(\mathrm{O_2})$$

Using this correction scheme, the DFT energy of $O_2$ was increased by 0.64 eV and the oxygen chemical potential was obtained as

$$\mu_{\mathrm{O}} = \frac{1}{2}E^{\mathrm{corr}}(\mathrm{O_2})$$

The carbon chemical potential was treated as a dependent quantity determined from the stability condition of $Li_2CO_3$ for given lithium and oxygen chemical potentials. This corresponds to evaluating $Li_2CO_3$ under the same chemical potential environment used for the voltage-dependent analysis, so that the voltage dependence of $\mu_{\mathrm{C}}$ arises from the $\mu_{\mathrm{Li}}(U)$ term:

$$2\mu_{\mathrm{Li}}(U) + 3\,\mu_{\mathrm{O}} + \mu_{\mathrm{C}} = E_{\mathrm{Li_2CO_3}}$$

$$\mu_{\mathrm{C}} = E_{\mathrm{Li_2CO_3}} - 2\mu_{\mathrm{Li}}(U) - 3\,\mu_{\mathrm{O}}$$

To evaluate the dependence of the fluorine substitution energy on the fluorine chemical environment, $F_2$ and $OF_2$ were additionally considered as alternative fluorine reference reservoirs. The corresponding fluorine chemical potentials were defined as

$$\mu_{\mathrm{F}}^{\mathrm{F_2}} = \frac{1}{2}E(\mathrm{F_2})$$

and

$$\mu_{\mathrm{F}}^{\mathrm{OF_2}} = \frac{1}{2}E(\mathrm{OF_2}) - \frac{1}{4}E(\mathrm{O_2})$$

In addition to the voltage-dependent chemical potentials, phase diagram based chemical potentials were also considered. The Li–C–O phase diagram is shown in **Figure S1**. The chemical potentials used for the voltage-dependent analysis and those obtained from the Li–C–O phase diagram are summarized in **Table S1(a)** and **Table S1(b)**, respectively. Conditions A–E denote selected phase equilibrium regions on the $Li_2CO_3$ stability boundary in the Li–C–O phase diagram.

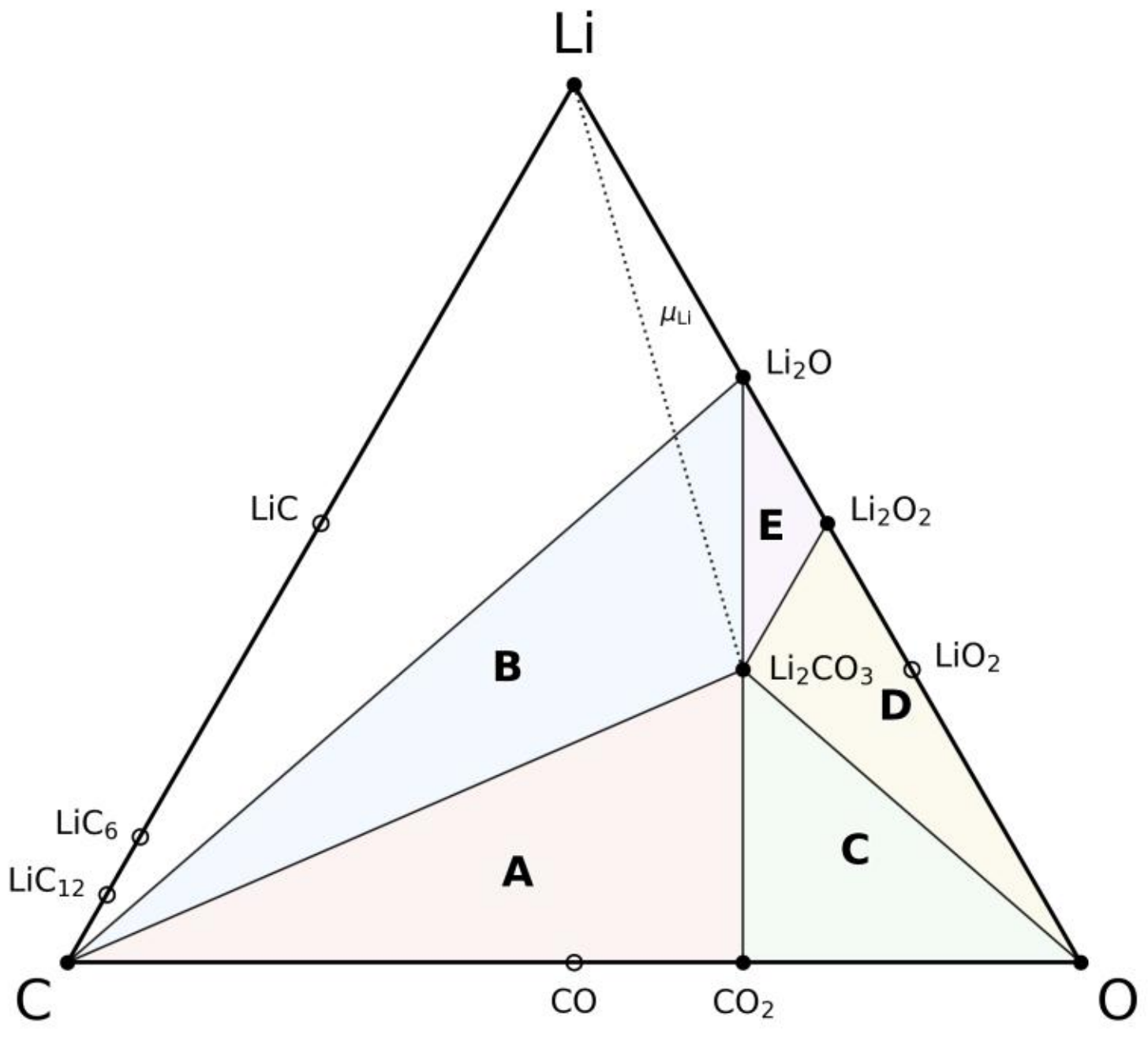

**Figure S1.** Schematic phase diagram of the Li–C–O system. The specific equilibrium phases corresponding to conditions **A–E** are listed in **Table S1(b)**.

**Table S1.** Chemical potential conditions used for defect formation energy calculations.

(a) Voltage-dependent chemical potential conditions used in the main analysis

| Voltage [V] | $\mu_{Li}$ [eV] | $\mu_{C}$ [eV] | $\mu_{O}$ [eV] |
|---|---|---|---|
| 0 | −2.01 | −24.27 | −7.51 |
| 2.96 | −4.97 | −18.35 | −7.51 |
| 3.82 | −5.83 | −16.63 | −7.51 |

(b) Phase-diagram based Li–C–O chemical potential conditions for additional calculations

| Condition | Equilibrium phases | $\mu_{Li}$ [eV] | $\mu_{C}$ [eV] | $\mu_{O}$ [eV] |
|---|---|---|---|---|
| A | $Li_2CO_3$–$CO_2$–C | −5.28 | −11.26 | −9.66 |
| B | $Li_2CO_3$–$Li_2O$–C | −3.37 | −11.26 | −10.93 |
| C | $Li_2CO_3$–$O_2$–$CO_2$ | −6.36 | −15.58 | −7.51 |
| D | $Li_2CO_3$–$O_2$–$Li_2O_2$ | −5.29 | −17.70 | −7.51 |
| E | $Li_2CO_3$–$Li_2O$–$Li_2O_2$ | −4.88 | −17.29 | −7.92 |

**S2. Crystal structural information of $Li_2O_2$ and $Li_2CO_3$**

**Table S2.** The Wyckoff positions and fractional coordinates of the optimized $Li_2CO_3$ structure. The corresponding experimental data are available in Ref. 4.

| Atom | Wyckoff site | *x* | *y* | *z* |
|---|---|---|---|---|
| Li | 8f | 0.1981 | 0.4484 | 0.8383 |
| C | 4e | 0 | 0.0660 | 0.25 |
| O(f) | 8f | 0.1474 | −0.0646 | 0.3146 |
| O(e) | 4e | 0 | 0.3223 | 0.25 |

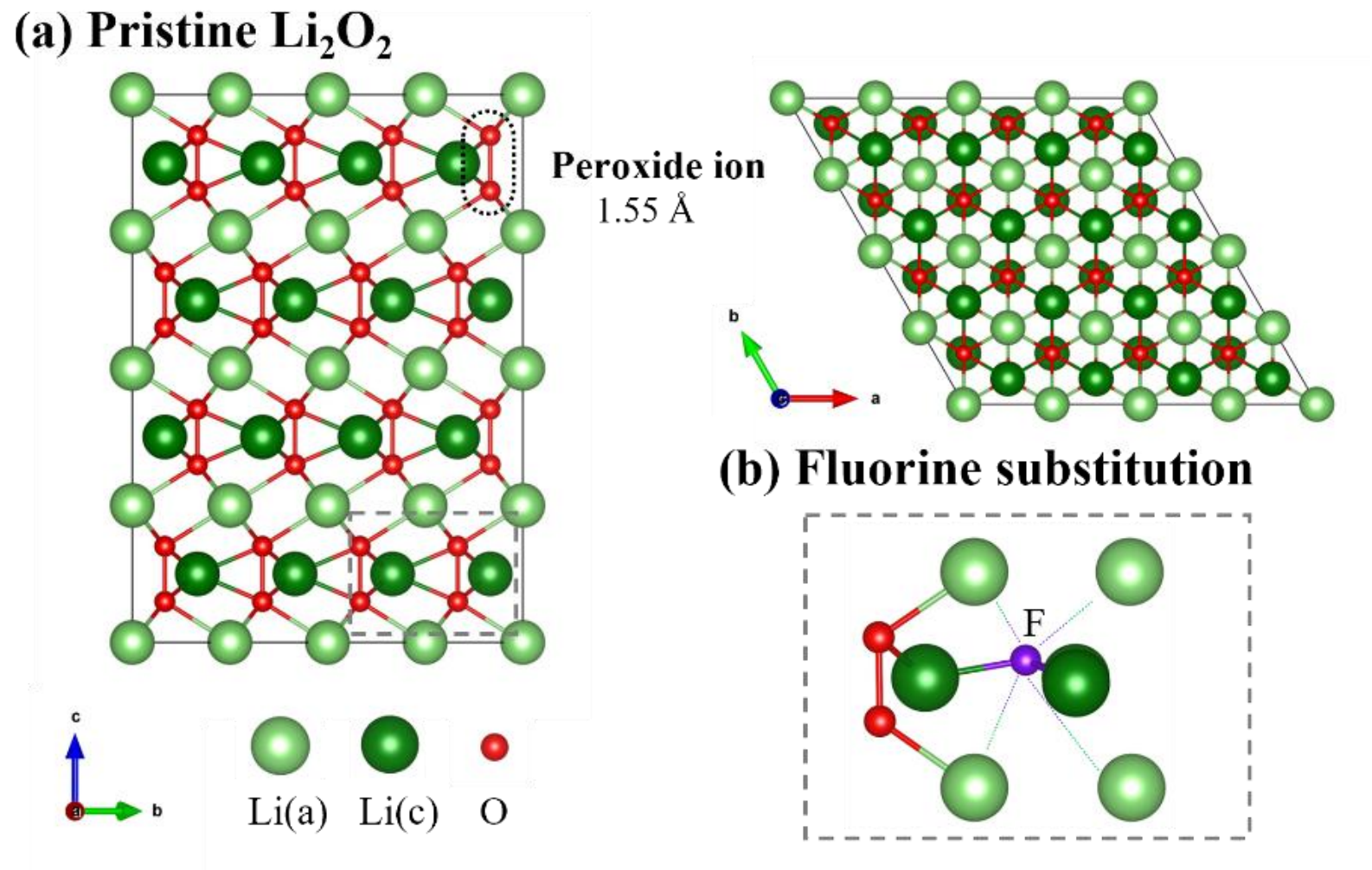

**Figure S2.** Crystal structure of $Li_2O_2$ with two inequivalent Li sites. Li(a) and Li(c) are assigned to octahedron and trigonal prism, respectively. (b) Optimized local structure of F substituted at peroxide ion. Dashed lines indicate the original Li–O coordination before F substitution.

## S3. Additional thermodynamic analyses

As a complementary reference to the voltage-dependent analysis in the main text, defect formation energies were calculated using the chemical potentials listed in **Table S1(b)**. The resulting values for conditions **A–E** are presented in **Figure S3**.

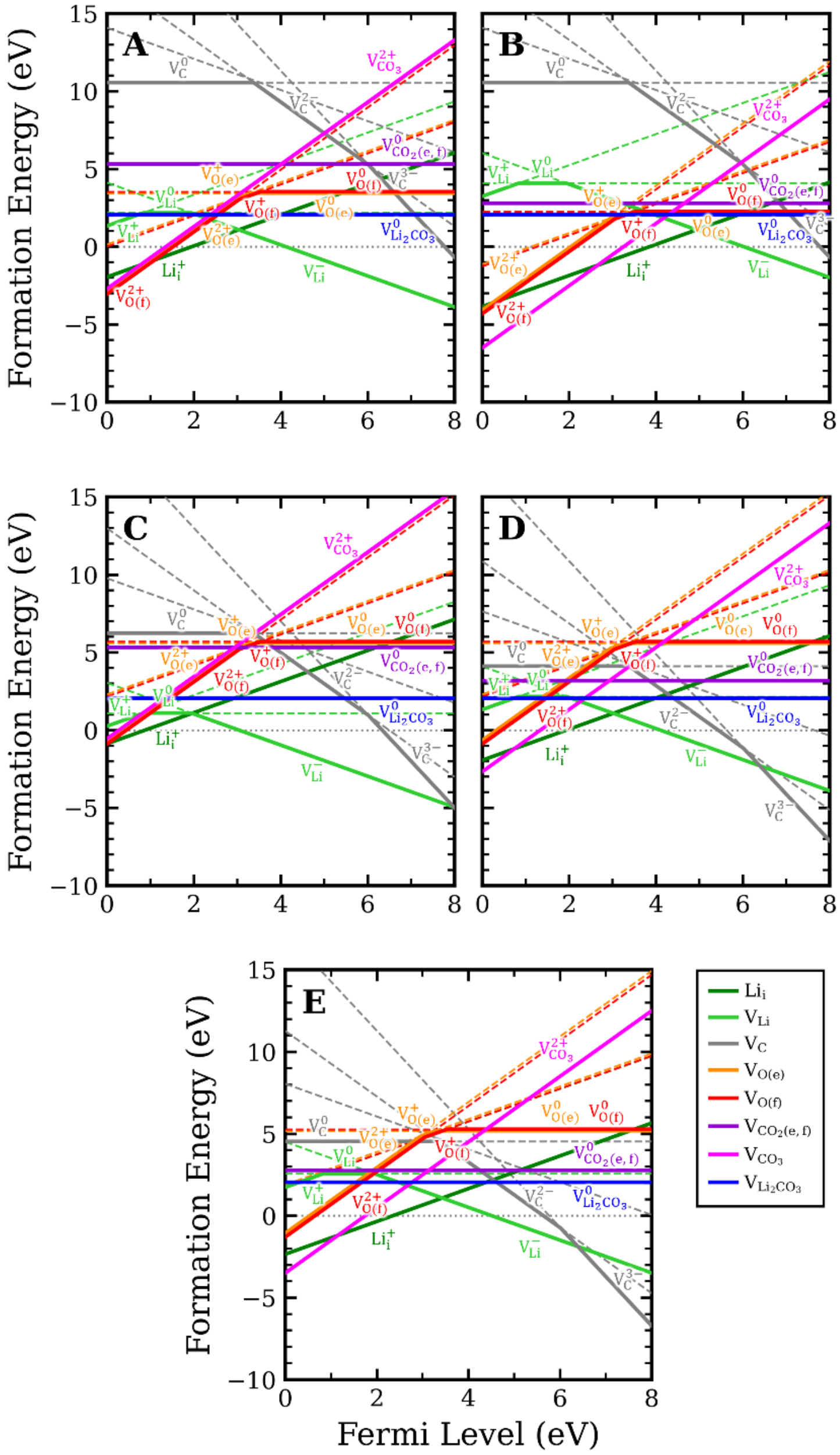

**Figure S3**. Defect formation energies calculated under conditions **A–E** based on the chemical potentials in **Table S1(b)**.

The fluorine substitution energies calculated using these reference reservoirs are summarized in **Table S3**. **Figure S4** summarizes the formation energies of charged vacancies separately for fluorine substitution at the O(f) and O(e) sites as a function of the Fermi level, using the chemical potentials corresponding to the equilibrium voltage of $Li_2CO_3$.

**Table S3.** Fluorine substitution energies calculated using different fluorine reference reservoirs. All energies are given in eV.

| Structure \ Reservoir | Min | | Max |
|---|---|---|---|
| | $F_2$ | $OF_2$ | LiF |
| $Li_2CO_3$, F at O(f) | 1.76 | 1.81 | 4.75 |
| $Li_2CO_3$, F at O(e) | 2.02 | 2.08 | 5.01 |
| $Li_2O_2$, F at $O_2$ dimer | −0.89 | −0.84 | 2.10 |

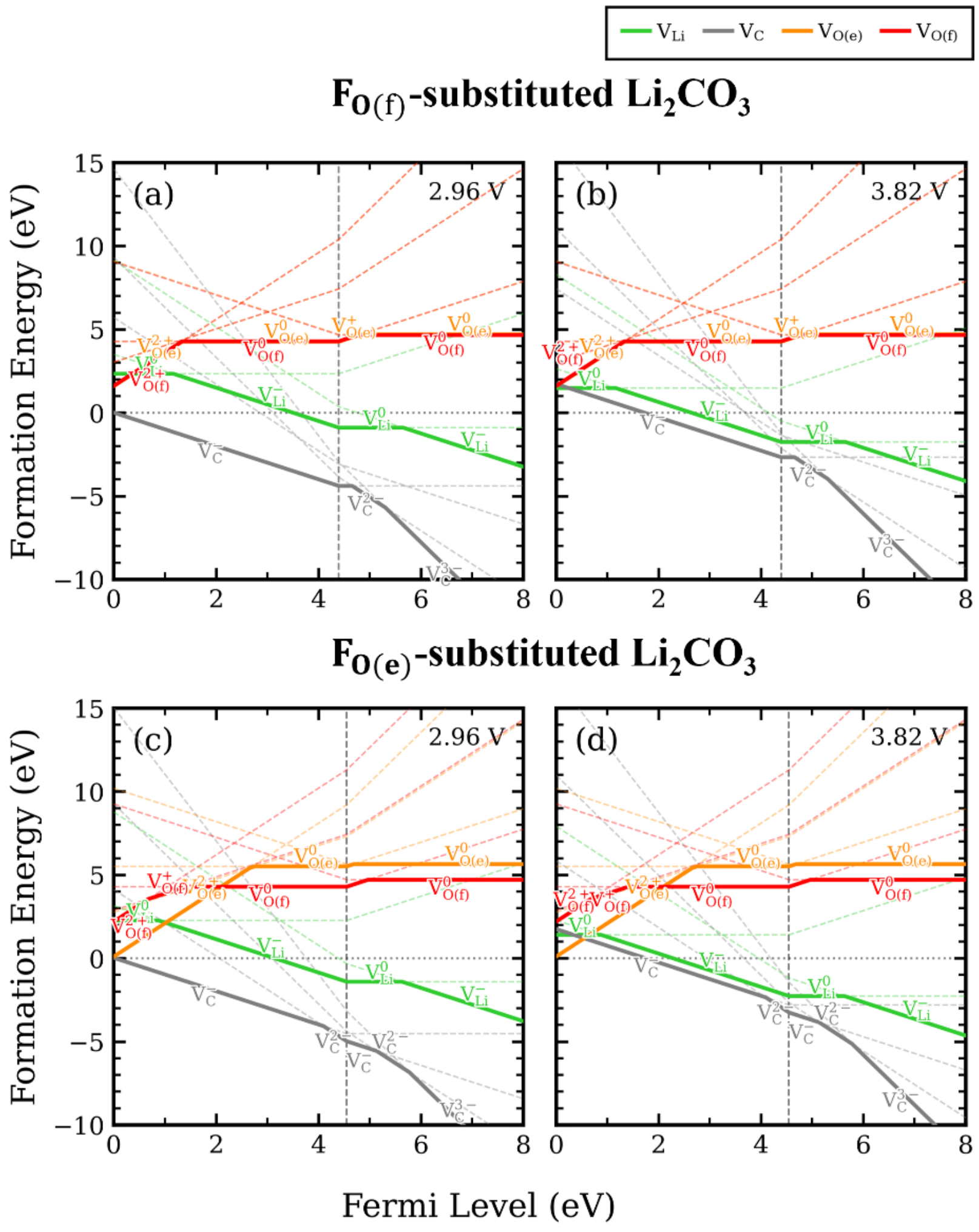

**Figure S4**. Vacancy formation energies in F-substituted $Li_2CO_3$. (a) and (b) show F substitution at the O(f) site, whereas (c) and (d) show F substitution at the O(e) site. The chemical potential conditions corresponding to 2.96 V are used in (a) and (c), and those corresponding to 3.82 V are used in (b) and (d). Formation energies are shown for Li, C, O(e) and O(f) vacancies with the thermodynamically stable charge state of the corresponding F-substituted structure used as the reference at each Fermi level. Gray dashed line indicates $F_O^+/F_O^0$ transition level. The Fermi level is referenced to the VBM of pristine $Li_2CO_3$. Solid lines indicate the lowest energy charge states, whereas dashed lines represent the higher energy charge states.

### S4. Structural and electronic changes in $Li_2CO_3$ induced by fluorine doping

For the structures in which fluorine substitutions at the two inequivalent oxygen sites, O(f) and O(e), the changes in local bond lengths are summarized in **Table S4**, while the corresponding charges in bond angles are shown in **Figure S5**. The local spin density distribution of the neutral fluorine substituted structures is presented in **Figure S6**, and the Bader charge analysis for atoms near the substituted fluorine site is shown in **Table S5**.

**Table S4**. Local bond lengths and their variation in pristine and F-doped structures

| Substituted structure | bond | Pristine length (Å) | F-doped length (Å) | Δ (Å) |
|---|---|---|---|---|
| $F_{O(f)}$-doped | C–F | 1.307 | 2.297 | +0.990 |
| | $C–O_1$ | 1.285 | 1.252 | -0.033 |
| | $C–O_2$ | 1.307 | 1.266 | -0.041 |
| | $Li_1–F$ | 1.964 | 1.813 | -0.151 |
| | $Li_2–F$ | 1.993 | 1.824 | -0.169 |
| | $Li_3–F$ | 2.038 | 1.835 | -0.203 |
| $F_{O(e)}$-doped | C–F | 1.285 | 2.150 | +0.865 |
| | $C–O_1$ | 1.307 | 1.264 | -0.043 |
| | $C–O_2$ | 1.307 | 1.264 | -0.043 |
| | $Li_1–F$ | 1.906 | 1.781 | -0.125 |
| | $Li_2–F$ | 1.906 | 1.781 | -0.125 |

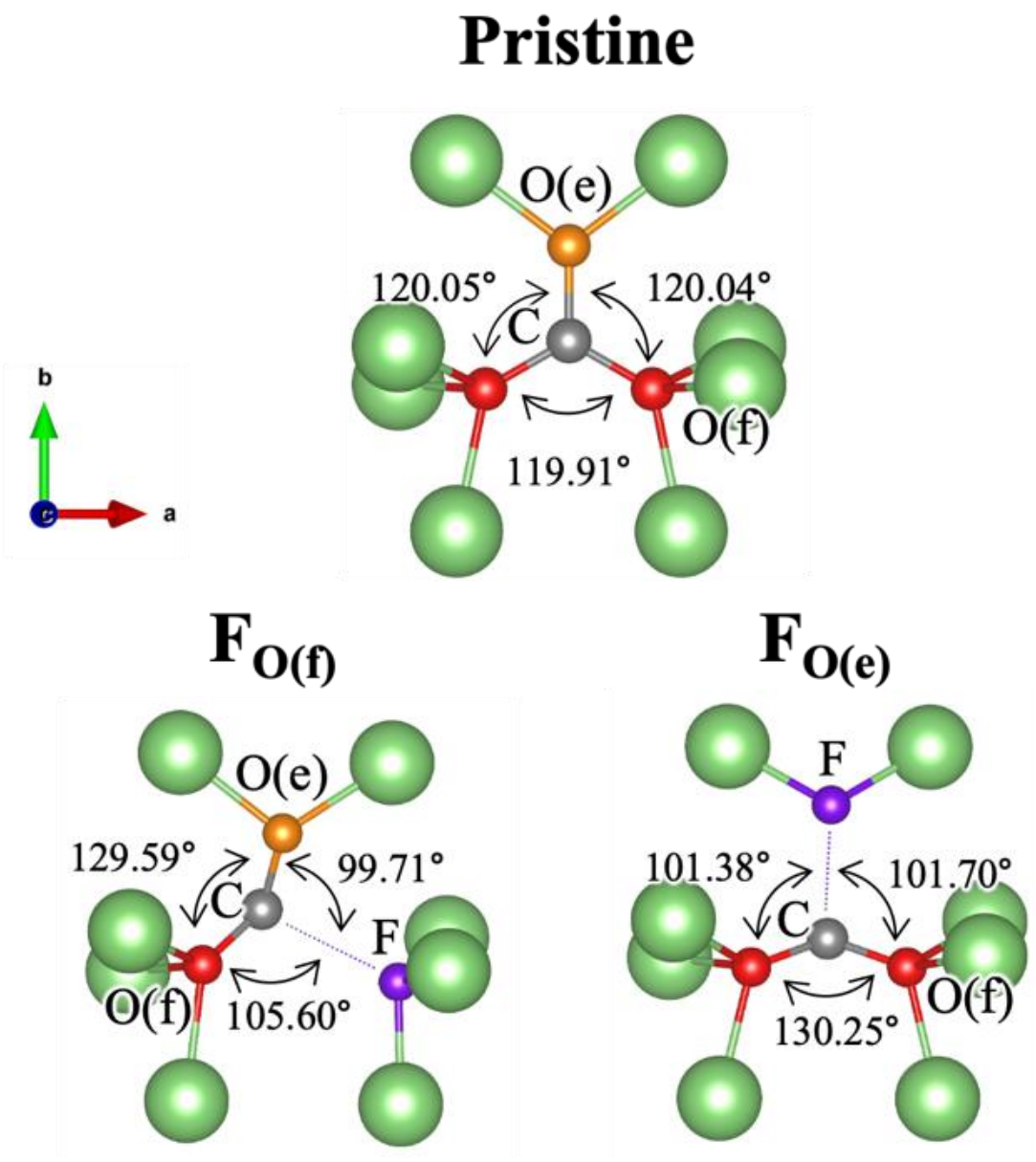

**Figure S5**. Bond angles in pristine and F-doped structures

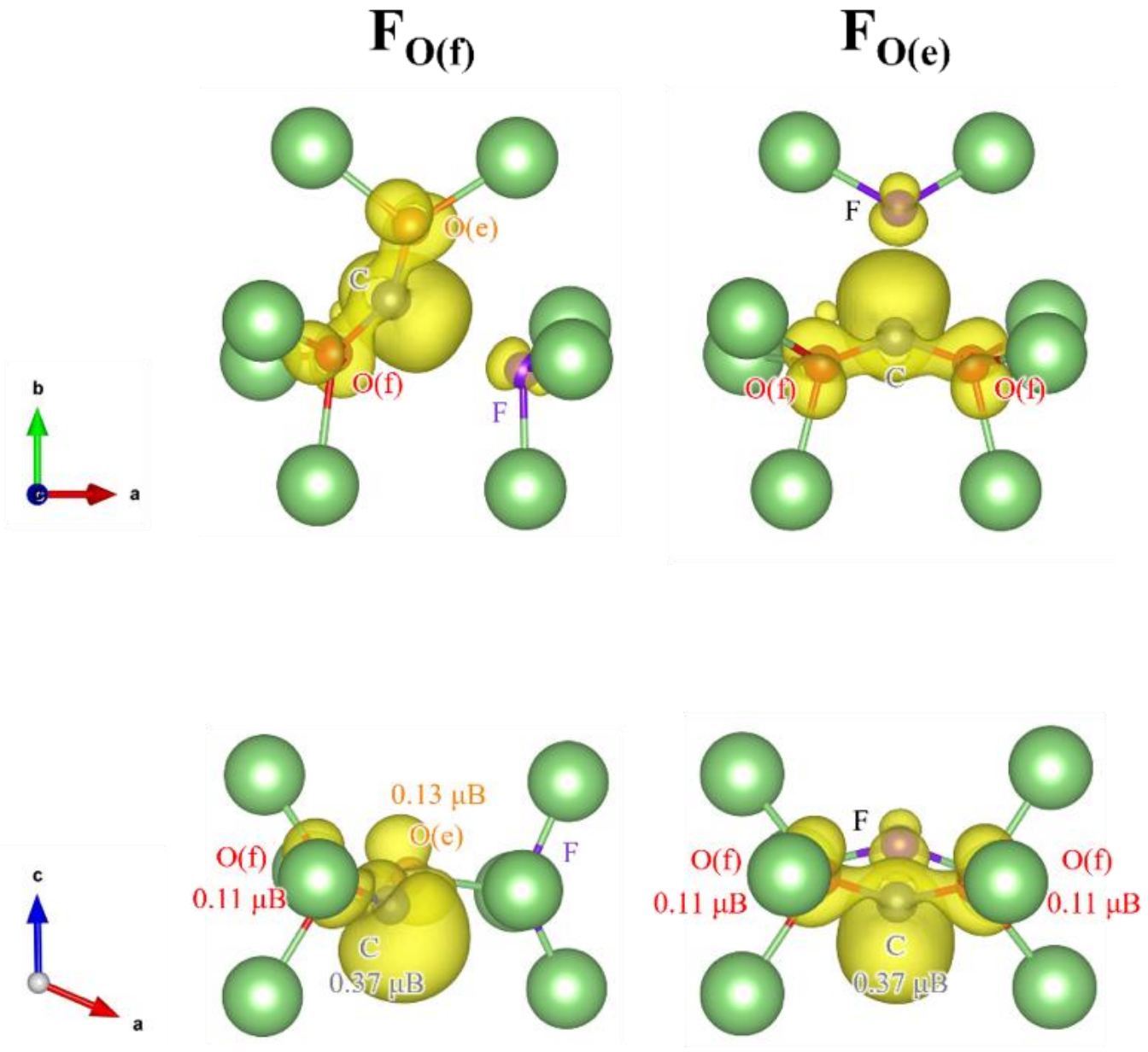

**Figure S6**. Spin density distribution around the neutral F-substitution defect in $Li_2CO_3$. The upper and lower rows show the $F_{O(f)}$ and $F_{O(e)}$ structures, respectively. The spin density is localized predominantly on the C atom adjacent to F and is partially distributed over the remaining O atoms of the same carbonate-derived unit.

**Table S5**. Bader charge analysis of pristine and F-doped structures

| Substituted structure | atom | Pristine $q$ (e) | F-doped $q$ (e) | $\Delta q$ (e) |
|---|---|---|---|---|
| $F_{O(f)}$-doped | substituted F | -1.387 | -0.880 | +0.507 |
| | C | +2.301 | +1.715 | -0.586 |
| | $O_1$ | -1.328 | -1.319 | +0.009 |
| | $O_2$ | -1.387 | -1.321 | +0.066 |
| | $Li_1$ | +0.901 | +0.899 | -0.002 |
| | $Li_2$ | +0.901 | +0.904 | +0.003 |
| | $Li_3$ | +0.901 | +0.903 | +0.002 |
| $F_{O(e)}$-doped | substituted F | -1.328 | -0.866 | +0.462 |
| | C | +2.301 | +1.730 | -0.571 |
| | $O_1$ | -1.387 | -1.346 | +0.041 |
| | $O_2$ | -1.387 | -1.334 | +0.053 |
| | $Li_1$ | +0.901 | +0.906 | +0.005 |
| | $Li_2$ | +0.901 | +0.905 | +0.004 |